# FPGA Implementation of an Intelligent Traffic Light Controller (I-TLC) in Verilog

1st Apoorva Banerjee
dept. of Electronics and Communication
Delhi Technological University
Delhi,India
apoorvabanerjee_me21b17_15@dtu.ac.in

*Abstract*—The objective of this paper is to design and implement an intelligent Traffic Light Controller (I-TLC) system for a four-way road intersection. The design is carried out using Verilog, and the hardware is implemented on a Field Programmable Gate Array (FPGA).The chosen intersection involves a 'main road' (heavy traffic flow) and a 'side road' (less traffic flow), which is equipped with sensors to detect the presence of traffic or pedestrians. The functionality of the system has undergone thorough verification through simulations conducted in the Xilinx ISE Design Studio software environment. Furthermore, it has been physically deployed on a Xilinx Spartan-3E FPGA board xc3s500e-4-fg320. A traffic light controller can be realized through the use of a microcontroller, Application-Specific Integrated Circuits (ASICs), or Field-Programmable Gate Arrays (FPGAs). FPGAs however offer significant advantages in terms of re-programmability, speed, and parallel processing capabilities, making them ideally suited for implementing complex, adaptive logic required by smart traffic management systems; thus, making this model of TLC extremely adaptive and cost efficient at the same time as compared to other existing models with reduced hardware usage and delay constraints.

## I. INTRODUCTION

In traditional traffic light controllers, an evident limitation is the absence of timing adjustability, meaning there's no capability to modify the traffic signal durations based on the vehicular density on the roads. Such an adaptable system plays a crucial role in alleviating the issue of severe traffic congestion in urban and suburban areas.Thus,there exists a demand for creating a rapid and secure traffic management system that can regulate vehicular flow during peak hours, eliminating the necessity for a traffic officer.TLC can be implemented using a microcontroller [14] [16], ASIC and FPGAs. Microcontrollers offer a flexible and cost-effective solution, allowing for easy programming and modification of the traffic light sequence and timing based on real-world requirements. They are ideal for small-scale or custom traffic management systems where adaptability is the key; ASICs are less flexible and more expensive to design and manufacture, making them suitable for large-scale deployments where the traffic control system does not require frequent updates whereas, Field Programmable Gate Arrays (FPGA) provide a more performance-oriented approach, enabling parallel processing capabilities that can manage complex traffic scenarios and integrate with sensors for adaptive traffic control in real-time. The main objective of this paper is to design an intelligent traffic controller with the help of timing mechanism and a sensor [8]. This document presents a design of an adaptive and an intelligent traffic light controller model, crafted using a Moore finite state machine.The design, which is programmed in Verilog, has been successfully implemented on an FPGA board Spartan-3E xc3s500e-4-fg320.The model proposed, demonstrates a very simple yet efficient approach to the of timing problems in the existing traffic light controller using minimal hardware utilization and minimised delay.

## II. RELATED WORK

Numerous studies have been conducted aiming to tackle complex issues within traffic control management. Yet, many of these attempts have not achieved efficient solutions, particularly regarding hardware utilization and timing constraints. The subsequent literature review highlights several of these research efforts from recent years.

In 2012, B. Dilip et al. [4] ,implemented a low cost advanced TLC system using Chip-Scope Pro and Virtual Input Output. The implemented Traffic Light Controller (TLC) encompassed four roads, along with sensor and camera support for the motorway. The system was realized in hardware through the use of a Spartan-3E FPGA.

In 2012, Sourav Nath et al. [11] designed and implemented an intelligent traffic light controller with VHDL and it was implemented on FPGA.A sensor and a camera for efficient monitoring had been incorporated the model, however the interfacing of the same was left as a future scope. The design was well-organised although nearly 6% of the LUTs available were utilized, which led to increased hardware utilization.

In the year 2015, Mr. Shashikant V. Lahade et al. [9] proposed a model of an intelligent and adaptive TLC system using Verilog. The model was implemented and the results were compared among Spartan 3E, Spartan 6 and IRTEX S. Though the model was quite methodical as it deploys an IR Sensor, which allows the system to operate efficiently during day and night. The only drawback in the model was the increased number of LUTs needed in Spartan 3E, which was approximately 8%.

In 2016, Ali K et al. [1] designed an intelligent traffic light control system using FPGA and VHDL. Furthermore, various functional layers were integrated, including the incorporation of a standby control signal, management of motion sensors, execution of special requests, and a design tailored for heavy traffic conditions. The functionality of these additions was confirmed and simulated with ModelSim.

In 2019, Nour T. Gadawe et al. [6] designed a smart traffic light controller using VHDL and implemented the same on Spartan 3E FPGA board. It was designed for a 4-road cross intersection, however it utilized 141 LUTs out of 9312, which led to increased hardware usage. However, the model's efficiency is highlighted by the incorporation of a walk latch circuit design, a feature that facilitates pedestrian road crossing when needed.

## III. FOUR-ROAD STRUCTURE

A standard traffic light controller is equipped with three lights: green signifies a 'proceed' signal, amber indicates a 'slow down to stop' signal, and red signifies 'stop.' The model presented in this paper of an intelligent traffic light controller has a sensor that senses the presence or absence of vehicles and/or pedestrians and reacts accordingly. A crossroad has been assumed wherein, a busy 'main road' or 'highway' is stretched out in the East-West direction and a less busy 'side road', which intersects it in a North-South direction as shown in Fig 1. [6] [19]. The primary goal in creating an intelligent and smart Traffic Light Controller (TLC) system is to minimize the waiting time for drivers at traffic lights on less busy side roads that experience little to no vehicular or pedestrian activity during peak hours of the day. On both the sides of the side road, i.e. North facing as well as South facing, a sensor 'C' has been placed [9] [2]. In the absence of any cars or pedestrians on the side road, the traffic lights on the main road stay 'Green,' prioritizing the flow of traffic on the main road. However, if a vehicle or pedestrian is detected on the side road by the sensor, the main road lights transition from 'Green' to 'Amber' and eventually to 'Red.' This shift allows the side road lights to turn 'Green.' The side road light remains 'Green' under two conditions: there is a vehicle on the side road, and the predefined duration of the green signal timer has not elapsed.

## IV. WORKING OF THE I-TLC SYSTEM

### A. WORKING PRINCIPLE

This paper introduces an interval timer system consisting of TS, which produces a shorter duration timing pulse, and TL, which generates a longer duration pulse, in response to a start timer (ST). TS is employed for timing the 'Amber' lights, while TL is utilized for timing the 'Green' lights. The Verilog code defines the input and output signals, as outlined in Tables 1 and 2.

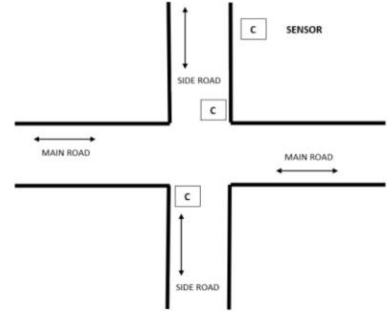

Fig. 1. RTL Schematic of the I-TLC system

TABLE I
INPUT VARIABLES

| Input Signal | Description |
|---|---|
| Reset | Returns the Finite State Machine (FSM) to its initial state |
| C | The sensor identifies the presence of vehicles, if any, on the side road. |
| TS | Short duration pulse |
| TL | Long duration pulse |

TABLE II
OUTPUT VARIABLES

| Output Signal | Description |
|---|---|
| MG, MY, MR | Assert green/Amber/red lights on Main Road |
| SG, SY, SR | Assert green/Amber/red on Side road |
| ST | Start timer-either a short or long interval |

### B. STATE DIAGRAM

In the Moore finite state machine model, the output is linked solely to the current state, in contrast to a Mealy Machine where the output is influenced by both the input and the current state. Therefore, for the Traffic Light Controller (TLC) system examined in this paper, four distinct scenarios, each representing a unique state of the FSM model, have been elucidated. Table 3 depicts the list of the following states along with its description.

TABLE III
STATE TABLE

| State | Description |
|---|---|
| S0 | Main road Green and Side road Red |
| S1 | Main road Yellow and Side road Red |
| S2 | Side road Green and Main road Red |
| S3 | Side road Yellow and Main road Red |

1) S0 is considered to be the default state or the initial state. It is defined as the state when main road light is 'Green' and the side road light is 'Red'. The FSM stays in the S0 state if TL has not expired yet or no vehicle is detected on the side road by the sensor.
2) While being in the S0 state, if both, TL has expired and a vehicle is also detected on the side road, a transition is made from state S0 to state S1. S1, as previously defined, is the state where the main road light is 'Amber' and the side road light is 'Red'. The timer ST is also made to restart the long timer TL. As TL expires, the system makes a change from state S0 to S1.
3) ST i.e the timer signal is made to start the short timer TS. The FSM remains in the state S1 until the short timer (TS) reaches its expiration. Upon TS expiration, the FSM transitions from state S1 to state S2, where the main road light is 'Red,' and the side road light is 'Green'.
4) When in state S2, the Finite State Machine (FSM) remains in that state as long as TL is active and there is still a vehicle on the side road. However, if TL has elapsed or there are no vehicles detected on the side road, the FSM transitions from S2 to S3, concurrently restarting the timer ST.
5) When in state S3, the traffic lights on the side road are Amber, and those on the main road are 'Red.' The FSM remains in state S3 until the short timer (TS) has not expired. However, if TS has elapsed, a transition occurs from S3 back to the initial state S0, with MG and SR both set to 1.

Fig 2 shows the state diagram of the I-TLC System.

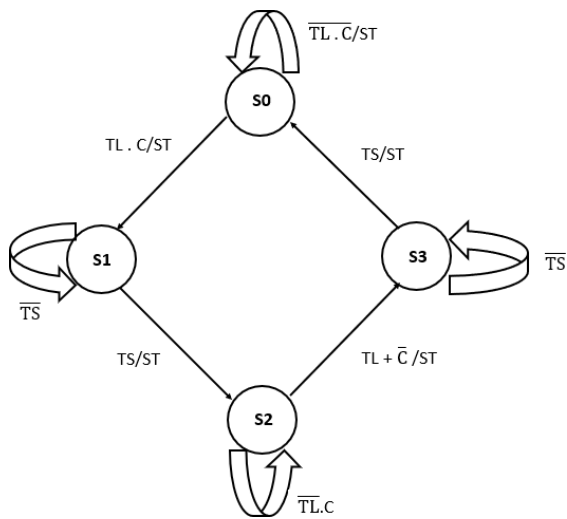

Fig. 2. State Diagram

## V. HARDWARE IMPLEMENTATION

The design of the traffic light controller system is realized by synthesizing the Verilog structural code design, followed by the generation of the bit file using Xilinx ISE 14.7 tools. Subsequently, this bit file is uploaded to the FPGA Spartan 3E development kit xc3s500e-4-fg320. Inputs to the FPGA board were provided manually viz, start timer (ST) was mapped to the push button K17 on the board; sensor (C), short timer (TS) and long timer (TL) were mapped to the slide switches N17,H18 and L14 respectively on the board. The system's outputs are displayed on the LEDs of the FPGA board with the MR,MY,MG being mapped to LEDs F9,E9,D11 respectively and SR,SY,SG being mapped to LEDs F11,E11 and E12 respectively on the FPGA board.

Fig 3 shows the implementation of the State S0 with MG=1, SR=1 and C=0 (no vehicle detected on the side road). The output is shown by the glowing LEDs D11 and F11.

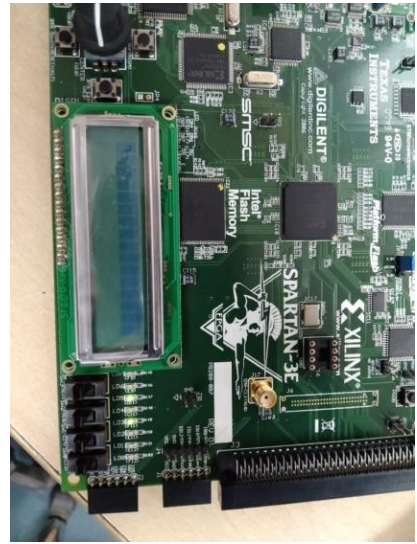

Fig. 3. State S0

Fig 4 shows the implementation of the transition from state S0 to state S1 as C=1, MY=1 and SR=1.Output is shown by the glowing LEDs E9 and F11.

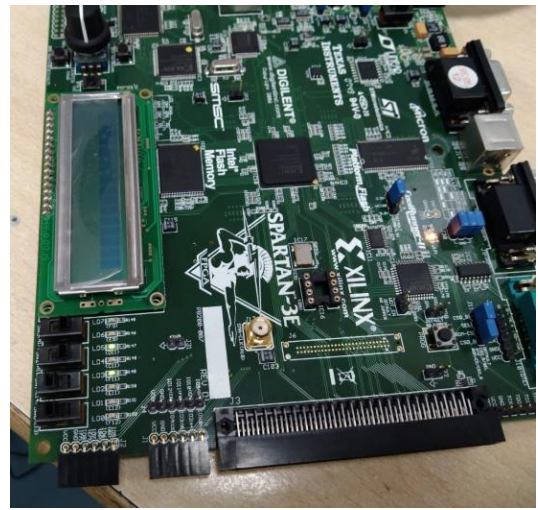

Fig. 4. State S1

Fig 5 shows the implementation of the transition from state S2 to S3 with MR=1,SG=1 and TL has expired i.e. TL=1.The output is displayed on the board by the glowing LEDs F9 and E11.

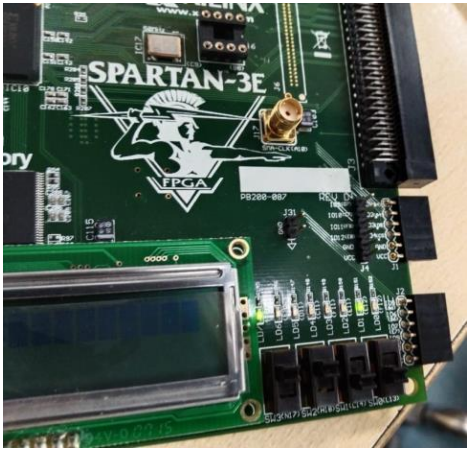

Fig. 5. State S2

Fig 6 shows the implementation of the state S3 with MR=1 and SY=1 with outputs being shown by the LEDs F9 and F11.

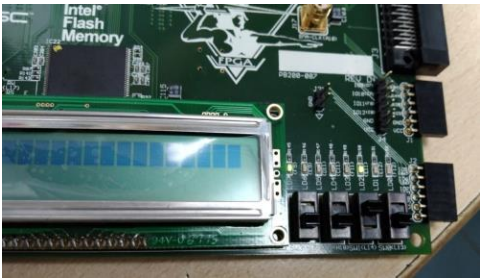

Fig. 6. State S3

## VI. SIMULATION AND RESULTS

Fig 7 illustrates the behavioral simulation at time t=29.5ns, depicting the state S0 with MG=1, SR=1, no car detected on the side road (sensor C=0), and both lights TL and TS not expired (TL, TS=0).

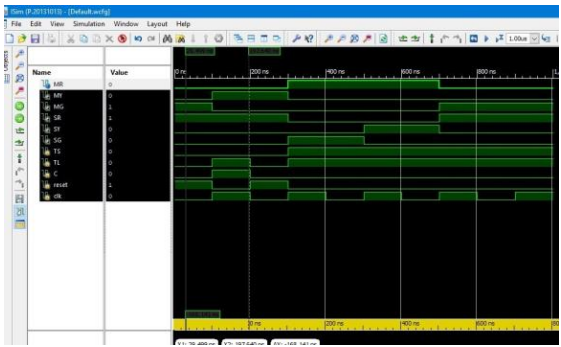

Fig. 7. State S0

Fig 8 illustrates the behavioral simulation at time t=144.5ns, showcasing the state S1, where MY=1, SR=1, the sensor detects a car on the side road (C=1), and the long timer TL has expired (TL=1).

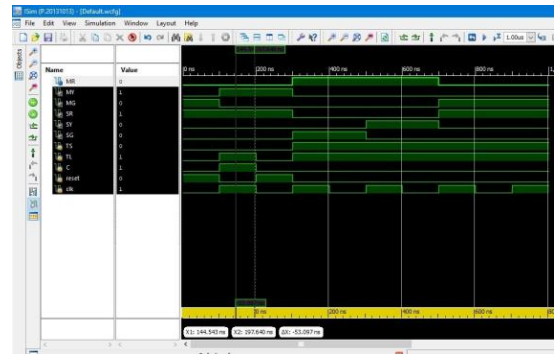

Fig. 8. State S1

Fig 9 displays the behavioral simulation at time t=374.631 ns, depicting the state S2 with MR=1 and SG=1. There is no vehicular movement detected on the side road, indicated by C=0, and the short timer has expired, resulting in TS=1.

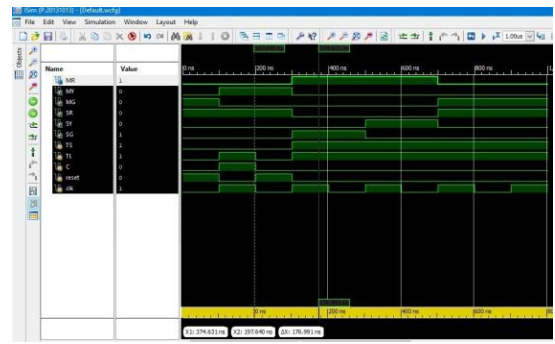

Fig. 9. State S2

Fig 10 illustrates the behavioral simulation at time t=566.3ns, portraying the state S3 with MR=1, SY=1, and both TS and TL set to 1.

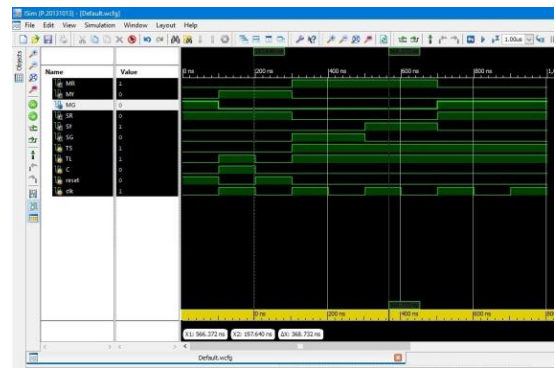

Fig. 10. State S3

Fig 11 displays the Register Transfer Level (RTL) schematic of the Intelligent Traffic Light Controller (I-TLC) system. The RTL schematic illustrates the logic implementation, providing a visualization of the data flow within the circuit.

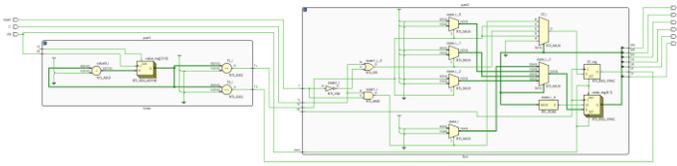

Fig. 11. RTL Schematic of the I-TLC system

The hardware resource utilization listed in Table 4.

TABLE IV
RESOURCE UTILIZATION

| Number of Slices | 4 out of 4656 |
|---|---|
| Number of SLice FLip FLops | 3 out of 9312 |
| Number of 4 input LUTs | 8 out of 931212 |
| Number of bonded IOBs | 12 out of 232 (5%) |
| Number of GCLKs | 1 out of 24(4%) |

## VII. CONCLUSION

The system has been modeled and simulated using Verilog to develop a Moore FSM model tailored for an Intelligent Traffic Light Controller (I-TLC) that dynamically adjusts its timing parameters in response to sensor data. Each component of the system was meticulously built and rigorously evaluated before proceeding to the next phase. The design process was thorough, with every design choice being carefully examined prior to implementation. The overall design and execution of the traffic light controller are commendable, paving the way for the development of more sophisticated systems. Remarkably, the design utilizes only 8 out of 931,212 Look-Up Tables (LUTs), a minimal amount when compared to some existing complex models. The maximum delay identified in the circuit is 5.982 nanoseconds. The Intelligent Adaptive Traffic Light Controller (IA-TLC) design is straightforward to implement across various FPGA boards, facilitating efficient traffic management.